\documentclass[twosided, 11pt]{article}
\usepackage{amssymb,amsmath}
\usepackage{mathabx}
\usepackage{times}
\usepackage{mathptmx}
\newtheorem{theorem}{Theorem}

\newtheorem{remark}[theorem]{Remark}
\newtheorem{definition}{Definition}

\newtheorem{example}{Example}

\usepackage[utf8]{inputenc}
\usepackage{enumitem}
\setlist{nosep} 

\newcommand\mrm{\mathrm}

\newcommand\acc[1]{\left\{#1\right\}}

\usepackage{xspace}
\usepackage{numprint}
\def\uint#1{\texttt{uint{#1}\_t}}
\def\ieme{\textsuperscript{th}\xspace}
\usepackage{todonotes}
\usepackage{marginnote}

\newcommand{\Ffour}{\textsf{F4}\xspace}
\newcommand{\Ffive}{\textsf{F5}\xspace}
\newcommand{\ie}{\mbox{\emph{i.e.}}\xspace}
\newcommand{\resp}{\mbox{resp.}\xspace}
\usepackage{url,amsmath}
\usepackage{booktabs}
\usepackage{multirow}

\newcommand{\gb}{Gr\"obner basis\xspace}
\newcommand{\gbla}{{\sffamily GBLA}\xspace}
\newcommand{\gblaone}{{\sffamily GBLA-v0.1}\xspace}
\newcommand{\gblatwo}{{\sffamily GBLA-v0.2}\xspace}
\newcommand{\lela}{{\sffamily LELA}\xspace}
\newcommand{\linbox}{{\sffamily LinBox}\xspace}
\newcommand{\fflas}{{\sffamily FFLAS-FFPACK}\xspace}
\newcommand{\magma}{{\sffamily Magma}\xspace}
\newcommand{\fl}{{\sffamily GB}\xspace}
\newcommand{\fli}{{\sffamily FL Implementation}\xspace}
\newcommand{\flr}{\fl reduction\xspace}
\newcommand{\openmp}{{\sffamily OpenMP}\xspace}
\newcommand{\omp}{{\sffamily OMP}\xspace}
\newcommand{\tbb}{{\sffamily Intel TBB}\xspace}
\newcommand{\starpu}{{\sffamily StarPU}\xspace}
\newcommand{\kaapi}{{\sffamily XKAAPI}\xspace}
\newcommand{\xk}{{\sffamily XK}\xspace}
\newcommand{\pthreads}{{\sffamily pthreads}\xspace}
\newcommand{\lp}{\ensuremath{\text{\tt lp}}}
\newcommand{\axpy}{{\tt AXPY}\xspace}
\newcommand{\trsm}{{\tt TRSM}\xspace}
\newcommand{\rk}{\ensuremath{\text{rank}}}

\newcommand{\npiv}{\ensuremath{n_\text{piv}}}
\newcommand{\pos}{\ensuremath{\text{\tt pos}}}
\newcommand{\val}{\ensuremath{\text{\tt val}}}
\newcommand{\ml}{\ensuremath{\text{\tt ml}}}
\newcommand{\dense}{\ensuremath{\text{dense}}}
\newcommand{\sizeof}{\ensuremath{\text{\tt sizeof}}}
\usepackage[normalem]{ulem}
\usepackage{algorithm}
\usepackage{algorithmicx}
\usepackage[noend]{algpseudocode}
\makeatletter
\renewcommand{\ALG@beginalgorithmic}{\small}
\makeatother
\usepackage{graphicx}
\usepackage{tikz}
\usetikzlibrary{arrows,matrix,fit,positioning}
\definecolor{mygreend}{HTML}{2ca92c}
\definecolor{myredd}{HTML}{c31313}
\def\ok{\bf}

\def\vs{vs.\xspace}

\begin{document}

%

\title{GBLA -- Gr\"obner Basis Linear Algebra Package}

\author{Brice Boyer$^1$, Christian Eder$^2$, Jean-Charles Faug\`ere$^1$,\\
Sylvian Lachartre$^3$, and Fayssal Martani$^4$\vspace{5mm}\\
$^1$INRIA, Paris-Rocquencourt Center, PolSys Project\\
UPMC, Univ. Paris 06, LIP6\\
CNRS, UMR 7606, LIP6\\
UFR Ing\'enierie 919, LIP6\\
Case 169, 4, Place Jussieu, F-75252 Paris\\
\texttt{brice.boyer@lip6.fr, jean-charles.faugere@inria.fr}\vspace{3mm}\\
$^2$University of Kaiserslautern\\
Department of Mathematics\\
PO box 3049\\
67653 Kaiserslautern\\
\texttt{ederc@mathematik.uni-kl.de}\vspace{3mm}\\
$^3$Thal\`es\\
\texttt{sylvian.lachartre@thalesgroup.com}\vspace{3mm}\\
$^4$\texttt{martani.net@gmail.com}
}

\maketitle
\begin{abstract}
This is a system paper about a new GPLv2 open source C library \gbla
implementing and improving the idea~\cite{FL-10b} of Faug\`ere and
Lachartre (\flr).
We further exploit underlying
structures in matrices generated during \gb computations in algorithms like
\Ffour or \Ffive taking advantage of block patterns by using a special data
structure called \emph{multilines}. Moreover, we discuss a new order of operations for the
reduction process. In various different experimental
results we show that \gbla performs better than \flr or \magma in sequential
computations (up to $40\%$ faster) and scales much better than \flr for a
higher number of cores: On $32$ cores we reach a scaling of up to $26$. \gbla
is up to $7$ times faster than \flr.
Further, we compare different parallel schedulers \gbla can be used with.
We also developed a new advanced storage format that exploits the fact that
our matrices are coming from \gb computations, shrinking storage by a factor
of up to $4$. A huge database of our matrices is freely available with \gbla.
\end{abstract}




\section{Introduction}\label{s:intro}
\begin{figure}
\begin{center}
\includegraphics[width=0.6\textwidth]{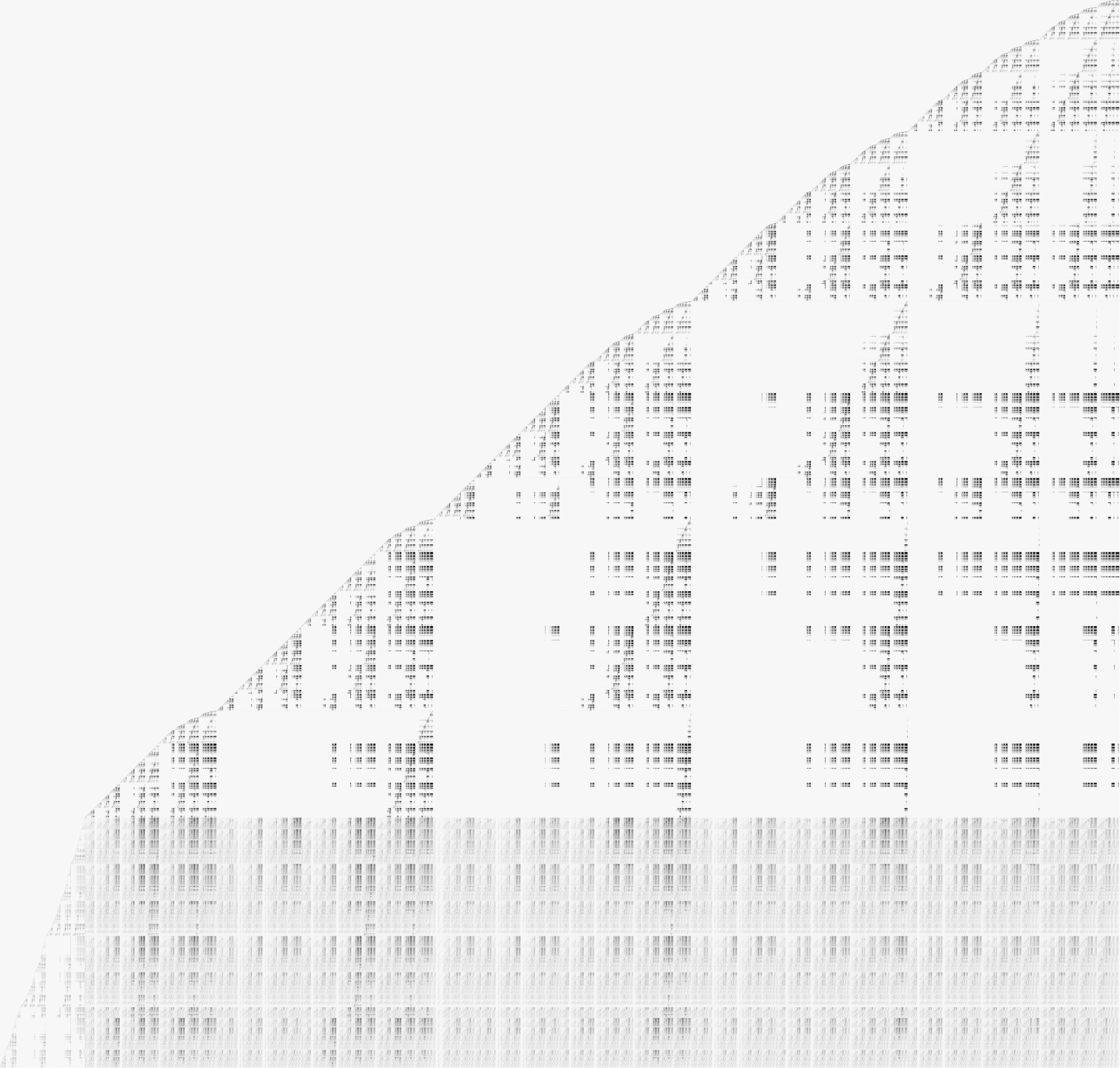}
\end{center}
\vspace*{-3mm}
\caption{\Ffour Matrix of degree $6$ for homogeneous
  \textsc{Katsura-12}}
\label{fig:input-matrix}
\end{figure}

In~\cite{Lach08,FL-10b}, Faug\`ere and Lachartre presented a specialized linear algebra
for \gb computation (\flr). The benefit of their approach is due to the
very special structure the corresponding matrices have. Using algorithms like
\Ffour the tasks of \emph{searching for reducers} and \emph{reducing the
input elements} are isolated. In the so-called \emph{symbolic preprocessing}
(see~\cite{F-F41999}) all possible reducers for all terms of a predefined subset
of currently available S-polynomials are collected. Out of this data a
matrix $M$ is
generated whose rows correspond to the coefficients of the polynomials
whereas the columns represent all appearing monomials sorted by the given
monomial order on the polynomial ring. New elements for the ongoing \gb computation
are computed via Gaussian elimination of $M$, \ie the reduction
process of several S-polynomials at once.
$M$ always has a structure like presented in
Figure~\ref{fig:input-matrix},
where black dots correspond to nonzero
coefficients of the associated polynomials. Faug\`ere and Lachar\-tre's idea is
to take advantage of $M$'s nearly in triangular
shape, already before starting the reduction process.

This is a system paper introducing in detail our new open source plain
C parallel library \gbla. This library includes efficient implementations not
only of the \flr but also new algorithmic improvements.
Here we present new ways of exploiting the underlying structure of the
matrices, introducing new matrix storage formats and various attempts to
improve the reduction process, especially for parallel computations.
We discuss different experimental results showing the benefits of our new attempt.

The paper is structured as follows:
In Section~\ref{sec:implementations} we give an overview of the structure of
our library.
Section~\ref{s:fmt} discusses the special matrix structure and presents a new efficient
storage format. This is important for testing and benchmarking purposes.
We also recall the general process used for reducing these matrices.
In Section~\ref{s:seq} we first review the main steps of the \fl
reduction. Afterwards we propose improvements to the sequential
algorithm by further exploiting patterns in the matrices. This is implemented in
our new library by specialized data structures and a rearrangement of the
order of steps of the \flr.
Section~\ref{s:par} is dedicated to ideas for efficient parallelization of our
library that also takes into account the improvements discussed beforehand.
In Section~\ref{sec:er} we show \gbla's efficiency by giving experimental results comparing
it to several other specialized linear algebra implementations for \gb
computations.

\section{Library \& Matrices}
\label{sec:implementations}
Our library is called \gbla (\gb linear algebra) and is the first plain
\textsl{C} open source package available for specialized linear algebra in
\gb-like computations. It is based on a first \textsl{C++} implementation of
Fayssal Martani in \lela, which is a fork of \linbox~\cite{DGGG+02} and which
is no longer actively developed.

The sources of our library are hosted at:
\url{http://hpac.imag.fr/gbla/}.
Under this website a database of our input matrices in different formats
(see Section~\ref{s:fmt}) is available as well as the routines
for converting matrices in our special format.

The general structure of the library is presented in
Table~\ref{tab:gbla-description}.

Input can come from files on disk or the standard input. The latter is
especially useful because we can use a pipe form \texttt{zcat} and never
uncompress the matrices to the disk. Uncompressed, our library of matrices
would reprensent hundreds of gigabytes of data.

\gbla supports the following data representations:
\begin{enumerate}
\item The code is optimized for prime fields
$\mathbb{F}_p$ with $p<2^{16}$ using SIMD vectorization~\cite{14BDG+}
and storing coefficients as \texttt{uint16}.
\item The library also supports a cofficient representation using \texttt{float}
such that we can use the better optimized SIMD instructions for floating point
arithmetic. $32$-bit floating points can be used for exact computations
over $\mathbb{F}_p$ with $p<2^{23}$.
\item There is also a version for $32$-bit field characteristic
using \texttt{uint32} data types that needs further optimization.
\end{enumerate}
\begin{table}
	\centering
  \scalebox{1}{
	\begin{tabular}{p{0.1\textwidth}|p{0.3\textwidth}|p{0.5\textwidth}}
		\toprule
    \multicolumn{1}{c|}{Folder} & \multicolumn{1}{c|}{Files} &
      \multicolumn{1}{c}{Description}\\
    \midrule
    \scriptsize{{\tt src}} & \scriptsize{{\tt types.*}} & general data types\\
    & \scriptsize{{\tt matrix.*}} & matrix and multiline vector types; conversion
    routines for sparse, hybrid, block, multiline matrices \\
    & \scriptsize{{\tt mapping.*}} & splicing of input matrix (Step~$1$ in \flr); different
    routines for usual block and multiline vector submatrix representations \\
    & \scriptsize{{\tt elimination.*}} & elimination routines including Steps~$2-4$ of \flr as
    well as adjusted routines for new order of operations (see
      Section~\ref{sec:new-order}) \\
    \cmidrule(lr){1-3}
    \scriptsize{{\tt cli}} & \scriptsize{{\tt io.*}} & input and output routines\\
    & \scriptsize{{\tt gbla.*}} & main routines for \flr \\
    \cmidrule(lr){1-3}
    \scriptsize{{\tt tools}} & \scriptsize{{\tt dump\_matrix.*}} & routines for
    dumping matrices; especially towards MatrixMarket or \magma formats\\
    & \scriptsize{{\tt converter.c}} & converting matrices from format 1 to
    format 2 (see Section~\ref{sec:file-format})\\
		\bottomrule
	\end{tabular}
  }
	\caption{Description of \gbla's structure}
	\label{tab:gbla-description}
\end{table}

Note that whereas it is true that vectorization in CPUs is faster for floating
point arithmetic compared to exact one we show in Section~\ref{sec:er} that for
$16$-bit computations memory usage can become a bottleneck:
Representing data via \texttt{uint16} can make matrices manageable that are not
feasible when using \texttt{float} data type.

For parallelization \gbla is based on \openmp. Current versions of \kaapi can
interpret \openmp macros, so one can also easily use \gbla with \kaapi as scheduler.

In order to assure cache locality we use blocks, all of them of dimension
$256 \times 256$ by default. The user has the freedom to set this to any
power of $2$, but in all of our experiments the preset size is advantageous due
to L1 cache size limitations.

At the moment we have two different types of implementations of the usual \flr (see
    Section~\ref{s:intro}) and the new order of operations for rank computations
(see Section~\ref{sec:new-order}) each:
\begin{enumerate}
\item The first type of implementations is completely based on the multiline data
structure, denoted \gblaone.
\item The second type is nearly always faster and denoted \gblatwo in this
paper. There we use multilines only in a very specific block
  situation where we can nearly guarantee in advance that they give a speedup
  due to cache locality. Otherwise we use
  usual sparse resp. dense row representations that are advantageous when
  sorting rows by pivots.
\end{enumerate}
Note that \gblatwo is able to reduce matrices that
  \gblaone cannot due to it's smaller memory footprint not introducing too many
  zeroes in multilines (see also Section~\ref{sec:er}).

%

\section{File formats and FL matrices}\label{s:fmt}
Input matrices in the \flr have some nearly-tri\-an\-gular structure
and patterns in the coefficients that we take advantage of. We describe \fl
matrices and an efficient way to store them.
\subsection{Description of FL matrices}
Matrices coming from \gb computations represent a set of polynomials in a polynomial
ring w.r.t. to some given monomial order. This order sorts the columns of the matrix,
each column represents a monomial. Each row represents a polynomial whereas the
entries are just the coefficients of the polynomial for the corresponding
monomial in the appropriate column. Due to this, \fl matrices are sparse. We can
assume that the matrix has been sorted by weight (number of non zero elements)
with row $0$ the heaviest\footnote{Throughout this paper, indexing is
zero-based.}.
%
Pivoting the rows corresponds to reordering of the polynomials; permuting non
pivot columns is allowed once before the \flr and re-done after the elimination
steps.
\par
The first non zero element on each row is a $1$ (each polynomial is monic), and
this element will be called \emph{pivoting candidate}. Every such pivoting
candidate lies below the main diagonal.
Columns whose last non zero element is not a pivoting candidate can be permuted
in order to separate them from the pivot ones.
\par
Now, the first $\npiv$ columns contain pivoting candidates, called
\emph{pivot columns}.  Among the pivoting candidates of a given column, one
row is selected, the \emph{pivot row}. This selection tries to keep $A$ (a
    $\npiv \times \npiv$ matrix) as sparse as
possible.
\subsection{Compressed binary format}
\label{sec:file-format}
%

In standard
Matrix Market\footnote{\url{http://math.nist.gov/MatrixMarket/}} file format,
\fl matrices are huge (hundreds of Gb) and slow to read. We
compress them to a CSR-like (Compressed Storage Row) format and store them in
\emph{binary} format (\ie streams of bytes rather than a text file).
We propose two different formats , see Table~\ref{tab:bin_formats}.
The files consist in consecutive sequences of elements of type \emph{size} repeated \emph{length} times.

\begin{table}
	\centering
  \scalebox{1}{
	\begin{tabular}{ccc ccc}
		\toprule
		\multicolumn{3}{c}{Format 1} & \multicolumn{3}{c}{Format 2} \\
		\cmidrule(lr){1-3}
		\cmidrule(lr){4-6}
		Size & Length & Data & Size & Length & Data \\
		\cmidrule(lr){1-3}
		\cmidrule(lr){4-6}
		          &     &      & \uint{32} & 1   & b      \\
		\uint{32} & 1   & m    & \uint{32} & 1   & m      \\
		\uint{32} & 1   & n    & \uint{32} & 1   & n      \\
		\uint{32} & 1   & p    & \uint{64} & 1   & p      \\
		\uint{64} & 1   & nnz  & \uint{64} & 1   & nnz    \\
		\uint{16} & nnz & data & \uint{32} & m   & rows   \\
		\uint{32} & nnz & cols & \uint{32} & m   & polmap \\
		\uint{32} & m   & rows & \uint{64} & 1   & k      \\
		          &     &      & \uint{64} & k   & colid  \\
		          &     &      & \uint{32} & 1   & pnb    \\
		          &     &      & \uint{64} & 1   & pnnz    \\
		          &     &      & \uint{32} & pnb & prow \\
			  &     &      & \texttt{xinty\_t} & pnnz& pdata  \\
			  \bottomrule
	\end{tabular}
  }
	\caption{Structure of the binary matrix formats.}
	\label{tab:bin_formats}
\end{table}

%
In Format~1, \texttt{m}, \texttt{n}, \texttt{p}, \texttt{nnz} \resp
represent the number of rows, columns, the modulo and the number of non zeros
in the sparse matrix. \texttt{rows[i]} represents the length of the $i$\ieme
row. If $j$ is the sum $\texttt{row[0]}+\dots+\texttt{row[i-1]}$, then on row
$i$, there is an element at column \texttt{cols[j+r]} with value
\texttt{data[j+r]} for all $r$ in $\acc{0,\dots,\texttt{row[i]}-1}$.
%
%

In Format~2, we separate the location of the non zero entries and the data.
We store the data of the polynomials separately since there is redundancy: many
lines will be of the form $m_i f_j$ where $m_i$ is some monomial and $f_j$ is a
polynomial in the intermediate \gb. Hence the coefficients in all lines of this type
correspond to the same polynomial $f_j$ and represent the same data, only the
location on the basis changes. We allow to store the data on
different machine types to adapt to the size of $p$. Data is
blocked, so we utilize the fact that several non zero elements on a row
may be adjacents, allowing compression of their consecutive column numbers.
In this format matrices must have less than $2^{31}$
rows.
\par
First, the lowest 3 bits of the first element \texttt{b} represent the value of
\texttt{x} and \texttt{y} in \texttt{xinty\_t}, namely $b=\texttt{u\ OR\ (v <<
1)}$ where $u$ is $1$ iff the type is signed and $y$ corresponds to a type on
$8 \cdot 2^v$ bits (for instance $\overline{011}^{2} = \texttt{1 OR (1 << 1)} $
represents \uint{16} type. On the highest bits a mask is used to store
a file format version.
\par
%
%
The $i\ieme$ row has
\texttt{rows[i]} elements. We prefer storing the row length since it fits on
$32$ bits while pointers (the accumulated row length) would fit on $64$ bits.
%
\par
%
%
We compress the column indices: If $s>1$ several non
zero elements are consecutive on a row and if $f$ is the first one, then we store
$f$ $s$ in the format.  If $s = 1$ then we use a mask and store $\texttt{f\
AND\ (1 << 31)}$. Here we lose a bit for the number of rows.
\par
So far, we have stored the locations of the non zero elements. The polynomial
data on a row is stored in \texttt{pdata} in the following fashion.
$\texttt{prow}[i]$ gives the $i\ieme$ polynomial number of elements (its support). There are
\texttt{pnb} polynomials.
%
%
$j=\texttt{polmap[i]}$ maps the polynomial number $j$ on row $i$.  The
polynomial data is laid out contiguously in \texttt{pdata}, polynomial $0$
finishes at $\texttt{pdata+prow[0]}$, polynomial $1$ finishes at
$\texttt{pdata+prow[0]+prow[1]}$, and so on.
%


In table~\ref{tab:compress} we show the raw size (in gigabits) of a few sparse
matrices in their binary format, compressed with
\textsf{gzip}\footnote{\url{http://www.gzip.org/}} (default options) and the
time it takes (in seconds). Compressing format 2 yields an $8$ time improvement
on the original uncompressed binary format 1 storage and over $4$ times better
than compressed format 1, in a much shorter time. The compressed format 2 is
hence much faster to load and it makes it easier to perform tests on.

\begin{table}
	\centering
  \setlength{\tabcolsep}{1.2pt}
  \scalebox{1}{
	\begin{tabular}{c| ccc ccc}
		\toprule
		\multirow{2}{*}{Matrix}& \multicolumn{3}{c}{Format 1} & \multicolumn{3}{c}{Format 2} \\
		\cmidrule(lr){2-4} \cmidrule(lr){5-7}
		& Size & Compressed & Time & Size & Compressed & Time  \\
		\midrule
		\texttt{F4-kat14-mat9}   &  2.3Gb    &   1.2Gb   &  230s    &  0.74Gb & 0.29Gb & 66s \\
		\texttt{F5-kat17-mat10}  &  43Gb     &   24Gb    &  4419s   &  12Gb   & 5.3Gb & 883s \\
		\bottomrule
	\end{tabular}
  }
	\caption{Storage and time efficiency of the new format}
	\label{tab:compress}
\end{table}

\section{Echelon Forms for Gr\"obner Bases}\label{s:seq}
In this section we present new developments in the implementation of the \flr
that can be found in our library (see Section~\ref{sec:implementations}).
Section~\ref{sec:multiline} presents ideas to exploit the structure of
the input \fl matrix $M$ further with dedicated data structures, and 
Section~\ref{sec:new-order} gives an alternative ordering of the steps of the
\flr if a non-reduced row echelon form of $M$ is sufficient.
\subsection{The reduction by Faug\`ere and Lachartre}
\label{sec:intro-fl}

There are $4$ main steps in the \flr:
\begin{enumerate}
\item
The input matrix $M$ already reveals a
lot of its pivots even before the Gaussian elimination starts.
For exploiting this fact we rearrange the rows and the
columns:
we reach a cutting of $M$ into
		$\begin{pmatrix}
			A & B \\
			C & D
		\end{pmatrix}$.
After this rearrangement one can see $4$ different parts of $M$: A very
sparse, upper triangular unit matrix on the top left ($A$) representing the already
known pivots. A denser, but still sparse top right part ($B$) of the same number of
rows. Moreover, there are two bottom parts, a left one which is still rather
sparse ($C$) and a right one, which tends to be denser ($D$). Whereas $A$
represents already known leading terms in the intermediate \gb, $D$
corresponds to the new polynomials added to the basis after the reduction
step. If $M$ is of dimensions $m \times n$ and if $\npiv$ denotes the
number of known pivots the characteristics of the four splices of $M$ are given
in Table~\ref{tab:fl-splices}.
\begin{table}
	\centering
  \scalebox{0.85}{
	\begin{tabular}{c|c|c}
		\toprule
    Splice & Dimensions & approx. density\\
    \cmidrule(lr){1-3}
    $A$ & $\npiv \times \npiv$ & $< 2\%$ \\
    $B$ & $\npiv \times (n - \npiv)$ & $12\%$ \\
    $C$ & $(m-\npiv) \times \npiv$ & $15\%$ \\
    $D$ & $(m-\npiv \times n-\npiv)$ & $35\%$\\
		\bottomrule
	\end{tabular}
  }
	\caption{Characteristics of matrix splices in \flr}
	\label{tab:fl-splices}
\end{table}
In general, $\npiv \gg m - \npiv$ and $\npiv \gg n - \npiv$.

\item In the second step of the \flr the known pivot rows are reduced with each
other, we perform an \trsm. Algebraically, this is equivalent to computing
$B \gets A^{-1} \times B$.
$A$ is invertible due to being upper triangular with $1$s on the diagonal. From
an implementational point of view one only reads $A$ and writes to $B$.
After this step, we receive a representation of $M$ in the format
		$\begin{pmatrix}
			\text{Id}_{\npiv} & A^{-1} \times B \\
			C & D
		\end{pmatrix}$.
\item In the third step, we reduce $C$ to zero using the identity matrix
$\text{Id}_{\npiv}$ performing \axpy. Doing this we also have to carry out
the corresponding operations induced by $B$ on $D$.
We get
		\[\begin{pmatrix}
			\text{Id}_{\npiv} & A^{-1} \times B \\
			0 & D - C \times \left(A^{-1} \times B\right)
		\end{pmatrix}.\]

\item The fourth step now reveals the data we are searching for: Via computing a
Gaussian Elimination on $D - C \times \left(A^{-1} \times B \right)$ we receive
new pivots reaching an upper triangular matrix $D'$. Those new pivots
correspond to new leading terms in our \gb, thus
the corresponding rows represent new polynomials to be added to the basis. On
the other hand, rows reducing to zero correspond to zero reductions in the \gb
computation.
\item As the last step we rearrange the columns of the echelon form of
$M$ and read off polynomials whose monomials are
sorted correctly w.r.t. the monomial order.
\end{enumerate}

If one is interested in a reduced row echelon form of $M$ we have to perform
the \flr a second time, but only on the right part $\begin{pmatrix} A^{-1} \times B \\
D'\end{pmatrix}$. From the \gb point of view a fully reduced row echelon form
has the advantage that also the multiples of polynomials already in the basis,
\ie elements representing the rows $\begin{pmatrix} A & B\end{pmatrix}$ are
reduced. Thus, reusing them in later reduction steps of the \gb computation can
be beneficial; we refer to Section~$2.4$
in~\cite{F-F41999} discussing the \texttt{Simplify} procedure.

\subsection{Multiline data structure}
\label{sec:multiline}
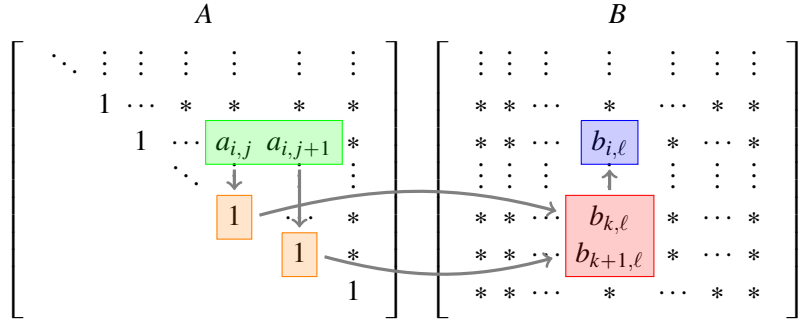
\begin{figure}
\centering
\begin{tikzpicture}[scale=0.9]
\begin{scope}[shift={(0,0)}]
  \matrix [matrix of math nodes, nodes={minimum height=10pt,text height=1.5ex,
    text depth=.3ex, transform shape}, nodes in empty cells,
  row sep=-2pt, column sep=-2pt, left delimiter={[},right delimiter={]}] (m)
  {
    \ddots &\vdots &\vdots &\vdots &\vdots &\vdots &\vdots\\
      & 1 & \cdots &* &* &* &* \\
      &  & 1 & \cdots & a_{i,j} & a_{i,j+1} &*\\
      &  &   & \ddots &\vdots & \vdots &\vdots\\
      &  &   &       & 1 & \cdots &*\\
      &  &   &       &   & 1 &* \\
      &  &   &       &   &   &1 \\
  };
  \node[above=1pt of m]{$A$};
  \fill[draw=green, fill=green, fill opacity=0.2] (m-3-5.north west) -- (m-3-6.north east) -- (m-3-6.south east) -- (m-3-5.south west) -- (m-3-5.north west);
  \fill[draw=orange, fill=orange, fill opacity=0.2] (m-5-5.north west) -- (m-5-5.north east) -- (m-5-5.south east) -- (m-5-5.south west) -- (m-5-5.north west);
  \fill[draw=orange, fill=orange, fill opacity=0.2] (m-6-6.north west) -- (m-6-6.north east) -- (m-6-6.south east) -- (m-6-6.south west) -- (m-6-6.north west);
  \draw[color=gray,very thick,<-,shorten >=2pt, shorten <=2pt](m-5-5.north) -- (m-3-5.south);
  \draw[color=gray,very thick,<-,shorten >=2pt, shorten <=2pt](m-6-6.north) -- (m-3-6.south);
\end{scope}
\begin{scope}[shift={(6.1,0)}]
  \matrix [matrix of math nodes, nodes={minimum height=10pt,text height=1.5ex,
    text depth=.3ex}, nodes in empty cells,
  row sep=-2pt, column sep=-2pt, left delimiter={[},right delimiter={]}] (n)
  {
    \vdots &\vdots&\vdots&\vdots&\vdots&\vdots&\vdots\\
    * &* &\cdots &* &\cdots &* &*\\
     * & * & \cdots &b_{i,\ell}&* &\cdots&* \\
    \vdots &\vdots&\vdots&\vdots&\vdots&\vdots&\vdots\\
     * & * & \cdots &b_{k,\ell}&* &\cdots&* \\
     * & * & \cdots &b_{k+1,\ell}&* &\cdots&* \\
    * &* &\cdots &* &\cdots &* &*\\
  };
  \node[above=1pt of n]{$B$};
  \fill[draw=blue, fill=blue, fill opacity=0.2] (n-3-4.north west) -- (n-3-4.north east) -- (n-3-4.south east) -- (n-3-4.south west) -- (n-3-4.north west);
  \fill[draw=red, fill=red, fill opacity=0.2] (n-5-4.north west)+(-0.2,0) rectangle (n-6-4.south east);
  \draw[color=gray,very thick,->,shorten >=2pt, shorten <=2pt](n-5-4.north) -- (n-3-4.south);

  \path[color=gray,very thick,->,shorten >=9pt, shorten <=3pt] (m-5-5.east) edge[out=15,in=165] (n-5-4.west);
  \path[color=gray,very thick,->,shorten >=5pt, shorten <=3pt] (m-6-6.east) edge[out=-15,in=-165] (n-6-4.west);
\end{scope}
\end{tikzpicture}
\caption{Exploiting horizontal and vertical patterns in the \trsm{} step.}
\label{fig:multiline-illustration}
\end{figure}

As already seen in Section~\ref{sec:intro-fl}, matrices coming from \gb
computations are structured in a way that can be exploited for a
specialized Gaussian Elimination. Furthermore, there are even more patterns in
such matrices that we use in order to speed up the computations. In
Figure~\ref{fig:input-matrix} we can see that the nonzero entries are, in
general, grouped in blocks.
In other words, if there is a nonzero element $m_{i,j}$ at position $j$
in row $i$ then also $m_{i,j+1}$ (horizontal pattern) and $m_{i+1,j}$ (vertical
pattern) tend to be nonzero, too. This fact can be used, for example, to optimize the
\axpy{} \resp{} \trsm{} computations in the second step of the \flr as
illustrated in Figure~\ref{fig:multiline-illustration}:
Assuming that $a_{i,j}$ and $a_{i+1,j}$ are both not zero (horizontal pattern),
element $b_{i,\ell}$ is updated by both nonzero elements $b_{k,\ell}$ and
$b_{k+1,\ell}$ (vertical pattern). Whereas the horizontal patterns are
canonically taken care ofstoring blocks row-wise, we have to pack
the vertical pattern in a dedicated data structure.

\begin{definition}
An \emph{$n$-multiline vector} $\ml$ is a data structure consisting
of two vectors in a sparse representation:
\begin{enumerate}
\item A position vector $\pos$ of column indices such that at each index at least one
of $n$ rows of elements has a nonzero element.
\item A value vector $\val$ of entries of $M$. The entries of all $n$ rows in
column $\pos[i]$ are stored consecutively, afterwards the $n$
entries at position $\pos[i+1]$ are stored. Note that $\val$ may have zero
elements.
\end{enumerate}
If $\pos$ has a length $\ell$, $\val$ has length $n \cdot \ell$. In this
situation we say that $\ml$ has length $\ell$.
For a $2$-multiline vector we use the shorthand notation \emph{multiline vector}.
\end{definition}

\begin{example}
Consider the following two rows:
\begin{center}
$
\begin{array}[]{rcccccccccc}
r_1 & = & [ & 2 & 0 & 0 & 1 & 0 & 0 & 5 & ],\\
r_2 & = & [ & 1 & 7 & 0 & 0 & 0 & 1 & 0 & ].
\end{array}
$
\end{center}
A sparse representation is given by $v_i$ (values) and $p_i$ (positions):
\begin{center}
$
\begin{array}[]{rcccccccccccccc}
v_1 & = & [ & 2 & 1 & 5 & ]& \;\; & v_2 & = & [ & 1 & 7 & 1 & ],\\
p_1 & = & [ & 0 & 3 & 6 & ]& \;\; & p_2 & = & [ & 0 & 1 & 5 & ].\\
\end{array}
$
\end{center}
A $2$-multiline vector representation of $r_1$ and $r_2$ is given by
\begin{center}
$
\begin{array}[]{lccccccccccccccc}
\ml.\val & = & [ & 2 & 1 & \color{myredd}{0} & 7 & 1 & \color{myredd}{0} &
\color{myredd}{0} & 1 & 5 & \color{myredd}{0} & ],\\
\ml.\pos & = & [ & 0 & 1 & 3 & 5 & 6 & ].\\
\end{array}
$
\end{center}
Four zero values are added to $\ml.\val$, two from $r_1$ and
$r_2$ resp. We do not add column $2$ since there
both, $r_1$ and $r_2$ have zero entries.
\end{example}

Multiline vectors are especially useful when performing \axpy{}. In the
following we use multiline vectors to illustrate the reduction of two
temporarily dense rows $\dense_1$ and $\dense_2$ with one multiline vector $\ml$
of length $\ell$. For the entries in $\ml.\val$ two situations are possible:
Either there is only one of $\ml.\val[2i]$ and $\ml.\val[2i+1]$ nonzero, or both are
nonzero. Due to the vertical pattern of \fl matrices very often both entries are
nonzero. We can perform a specialized
$\axpy$ operation on $\dense_1$ and $\dense_2$ with scalars $\lambda_{1,1},
\lambda_{1,2}$ coming from column $j$ and $\lambda_{2,1}, \lambda_{2,2}$
from column $j+1$ where $j$ is the loop step in the corresponding $\trsm$ opteration:

\begin{algorithm}
\begin{algorithmic}[1]
\Require $\dense_1$, $\dense_2$, $\lambda_{1,1}$, $\lambda_{1,2}$,
  $\lambda_{2,1}$, $\lambda_{2,2}$,
  $\ml$.
\State $v_1$, $v_2$, $i$, $k$
\For{$\left(i=0;\; i<\ell;\; i\gets i+1\right)$}
  \State $k \gets \ml.\pos[i]$
  \State $v_1 \gets \ml.\val[2i]$
  \State $v_2 \gets \ml.\val[2i+1]$
  \State $\dense_1[k] \gets \lambda_{1,1} v_1 + \lambda_{1,2} v_2$
  \State $\dense_2[k] \gets \lambda_{2,1} v_1 + \lambda_{2,2} v_2$
\EndFor
\end{algorithmic}
\caption{$\axpy$ of two dense rows of length $\ell$ with a multiline vector.}
\label{alg:ml-axpy}
\end{algorithm}

The benefit of Algorithm~\ref{alg:ml-axpy} is clear: We perform $4$ reductions
(each dense row is reduced by two rows) in one step. On the other hand, if the
horizontal pattern does not lead to two successive nonzero entries (for example
if $a_{i,j+1}$ is zero in Figure~\ref{fig:multiline-illustration}),
then Algorithm~\ref{alg:ml-axpy} would not use $\ml.\val[2i+1]$. This would
introduce an disadvantage due to using only every other element of $\ml.\val$. In our
implementation we take care of this situation and have a specialized $\axpy$
implementation for that. Still, we are performing two
reductions (each dense row is reduced by one row) in one step.

Assuming general $n$-multiline vectors the problem of introducing useless
operations on zero elements appears. For multiline vectors, \ie $n=2$, we can
perform lightweight tests before the actual loop to ensure execution only on
nonzero $\lambda_{1,1}, \lambda_{1,2}$ (for single $\axpy$) \resp
$\lambda_{1,1}, \lambda_{1,2}, \lambda_{2,1}, \lambda_{2,2}$ (for
  Algorithm~\ref{alg:ml-axpy}). For general $n$ we cannot predict every possible
configuration of the $n^2$ scalars $\lambda_{1,1},\ldots,\lambda_{n,n}$.
Moreover, for $n$-multiline vectors the memory overhead can get problematic,
too. For $n=2$ we can lose at most $\sizeof(\text{entry})$
bytes per column index, but for arbitrary $n$ this increases to $(n-1) \cdot
\sizeof(\text{entry})$ bytes.
All in all, we note the following fact that is also based on practical
experimental results.

\begin{remark}
Based on cache efficiency as well as memory overhead due to adding zero entries
to the \val{} vector $2$-multiline vector data structures are
the most efficient.
\end{remark}

As already mentioned in~\cite{FaLa10}, representing the matrices $A$, $B$, $C$ and $D$
in blocks has several benefits: Firstly, we can pack data in small blocks that
fit into cache and thus we increase spatial and temporal locality. Secondly,
separating the data into column blocks we can perform operations on $B$ and $D$
rather naturally in parallel. Thus we are
combining the multiline vector data structure with a block representation in our
implementation.
In the following, presented pseudo code is independent of the corresponding
row \resp block representation, standard row representation is used. Multiline
representations impede the readability of the algorithms, if there is an impact
on switching to multilines, we point this out in the text.

Using multilines is useful in situations where we can predict
horizontal \emph{and} vertical patterns with a high probability, in order to see
advantages and drawbacks we have two different implementations, \gblaone and
\gblatwo, which use multilines in different ways (see also
Section~\ref{sec:implementations}).

\subsection{New order of operations}
\label{sec:new-order}
If the number of initially known pivots (\ie the number of rows of $A$ and $B$)
is large compared to the number of rows of $C$ and $D$, then most work of the
\flr is spent in reducing $A$, the \trsm step $A^{-1} B$. For the \gb the new
information for updating the basis is strictly in $D$. Thus, if we are not
required to compute a reduced echelon form of the input matrix $M$, but if we
are only interested in the reduction of $D$ \resp the rank of $M$ we can omit
the \trsm step. Whereas in~\cite{FaLa10} the original \flr removes
nonzero entries above ($A^{-1}B$) and below (deleting $C$) the
known pivots, it is enough to reduce elements
below the pivots.

Thus, after splicing the input matrix $M$ of dimension $m \times n$ we can directly
reduce $C$ with $A$ while
reflecting the corresponding operations with $B$ on $D$ with the following steps.

\begin{algorithm}
\begin{algorithmic}[1]
\Require submatrices $A \left(\npiv \times \npiv\right)$, $B \left(\npiv \times
    (n-\npiv)\right)$, $C
    \left((m-\npiv) \times \npiv\right)$, $D \left((m-\npiv) \times
        (n-\npiv)\right)$.
\State $\dense_C$, $\dense_D$, $i$, $j$
\For{$\left(i=0;\; i<m-\npiv;\; i\gets i+1\right)$}
  \State $\dense_C \gets \textrm{copy\_sparse\_row\_to\_dense}(C[i,*])$
  \State $\dense_D \gets \textrm{copy\_sparse\_row\_to\_dense}(D[i,*])$
  \For{$\left(j=0;\; j<\npiv;\; j\gets j+1\right)$}
    \If{$\left(\dense_C[j] \neq 0\right)$}
      \State $\axpy\left(\dense_C, \dense_C[j], A[j,*]\right)$
      \State $\axpy\left(\dense_D, \dense_C[j], B[j,*]\right)$
    \EndIf
  \EndFor
  \State $D[i,*] \gets \textrm{copy\_dense\_row\_to\_sparse}(\dense_D)$
\EndFor
\end{algorithmic}
\caption{Reduction of $C$ and $D$}
\label{alg:new-red-order}
\end{algorithm}

Whereas Algorithm~\ref{alg:new-red-order} describes the idea of reducing $C$ and
$D$ from a mathematical point of view, in practice one would want to use a block
representation for the data in order to improve cache locality and also
parallelization. Strangely, it turned out that this is not optimal for efficient
computations: In order for a block representation to make sense one needs to
completely reduce all rows \resp multilines in a given block before reducing the
next block. That is not a problem for $B$ and $D$ since their blocks do not
depend on the columns, but it is disadvantageous for $A$ and $C$. Assuming an
operation on a lefthand side block of $C$ due to a reduction from a block from
$A$. Any row operation on $C$ must be carried out through all blocks on the
right. Even worse, whenever we would try to handle $C$ per row  \resp multiline
and $D$ per block at the same time this would introduce a lot of writing to $D$.
Thus, in our implementation we found the most efficient solution to be the
following:
\begin{enumerate}
\item Store $A$ and $C$ in multiline representation and $B$ and $D$ in block
multiline representation as defined in Section~\ref{sec:multiline}.
\item Carry out the reduction of $C$ by $A$, but store the
corresponding coefficients needed for the reduction of $D$ by $B$ later on.
\item Transform $C$ to block multiline
representation $C'$.
\item Reduce $D$ by $B$ using thecoefficients stored in
$C'$.
\end{enumerate}

Thus we have an optimal reduction of $C$ and an optimal reduction of $D$. The
only overhead we have to pay for this is the transformation from $C$ to $C'$.
But copying $C$ into block format is negligible compared to the reduction
operations done in $C$ and $D$.

In Section~\ref{sec:er} we see that this new order of operations is faster
than the standard \flr for full rank matrices from \Ffive{}. The density of
the row echelon form of $M$ does not vary too much
from $M$'s initial density which leads in less memory footprint.

\subsection{Modified structured Gaussian Elimination}
\label{sec:sge}
Computing the row echelon form of $D$ the original \fli used a sequential structured Gaussian Elimination. Here
we use a modified variant that can be easily parallelized.

\begin{algorithm}
\begin{algorithmic}[1]
\Require submatrix $D \left((m-\npiv) \times
        (n-\npiv)\right)$.
\Ensure $\rk_D$, rank of $D$
\State $\dense_D$, $i$, $j$
\State $\rk_D \gets 0$
\For{$\left(i=0;\; i<m-\npiv;\; i\gets i+1\right)$}
  \State $\text{normalize}(D[i,*])$
  \State $\dense_D \gets \textrm{copy\_sparse\_row\_to\_dense}(D[i,*])$
  \For{$\left(j=0;\; j<i-1;\; j\gets j+1\right)$}
    \If{$\left(\text{head}(\dense_D) = \text{head}(D[j,*])\right)$}
      \State $\axpy\left(\dense_D, \text{head}(\dense_D), D[j,*]\right)$
    \EndIf
  \EndFor
  \State $D[i,*] \gets \textrm{copy\_dense\_row\_to\_sparse}(\dense_D)$
  \State $\text{normalize}(D[i,*])$
  \If{$\left(\text{not\_empty}(D[i,*])\right)$}
    \State $\rk_D \gets \rk_D + 1$
  \EndIf
\EndFor
\State \Return $\rk_D$
\end{algorithmic}
\caption{Modified structured Gaussian Elimination of $D$}
\label{alg:sge}
\end{algorithm}

In Algorithm~\ref{alg:sge} we do a structured Gaussian Elimination on the rows
of $D$. Note that $D$ is not a unitary matrix, so normalization and inversions
are required. At the very end the rank of $D$ is returned. The modification lies
mainly in the fact that once we have found a new pivot row, we do not sort the
list of known pivot rows, but just add the new one. This is due to the usage of
multilines in our implementation. Storing two (or more) rows in this packed
format it is
inefficient to sort pivots by column index. Possibly we would need to open a
multiline row and move the second row to another position. For this, all
intermediate multiline rows need to be recalculated. Thus we decided to
relinquish the sorting at this point of the computation and sort later on when
reconstructing the row echelon form of the initial matrix $M$. Note that whereas
we use multilines everywhere in \gblaone, in \gblatwo (see Section~\ref{sec:er})
we restrict the usage of multilines to
specific block situations and no longer use them for the dense Gaussian
Elimination of $D$. Thus we are able to perform a sorting of the pivots.

\section{Parallelization}\label{s:par}
In this section we discuss improvements concerning parallelizing the \flr taking
the new ideas presented in the last section into account.
For this we have experimented with different parallel schedulers such as
\openmp, \kaapi and \pthreads. Moreover, whereas the initial implementation of
Faug\`ere and Lachartre used a sequential Gaussian Elimination of $D$ we are now
able to use a parallel version of Algorithm~\ref{alg:sge}.

\subsection{Parallelization of the modified structured Gaussian Elimination}
\label{sec:psge}
As already discussed in Section~\ref{sec:sge} we use a modified structured
Gaussian Elimination for multilines which omits sorting the list of known
pivots, postponed to the reconstruction of the echelon form of the input
matrix $M$. In our library \gbla there is also a non-multiline version with
sorting, see Section~\ref{sec:er} for more information.

Assuming that we have already found $k$ pivots in Algorithm~\ref{alg:sge}, we
are able to reduce several rows of index $>k$ in parallel. The $k$ pivots are
already in their normalized form, they are readonly, thus we can easily update
$
  D[\ell,*] \gets D[\ell,*] + \sum_{i=0}^k \lambda_i D[i,*]
$
for all $k < \ell < m -\npiv$ and corresponding multiples $\lambda_i$. Clearly,
this introduces some bookkeeping: Whereas in the above situation $D[k+1,*]$ is
fully reduced with the $k$ known pivots, the rows $D[k+j,*]$ for $j>1$ are not.
Thus we can add $D[k+1,*]$ to the list of known pivots, but not $D[k+j,*]$. We
handle this by using a global waiting list $W$ which keeps the rows not fully
reduced and the indices of the last pivot row up to which we have already
updated the corresponding row. Different threads share a global variable
$\lp$: the last known pivot. Each thread performs the following
operations:
\begin{enumerate}
\item\label{step:fetch}
Fetch the next available row $D[j,*]$ which was not updated up to this
point or which is already in the waiting list $W$.
\item Reduce it with all pivots not applied until now, up to $\lp$.
\item\label{step:add-piv}
If $j=\lp +1$, $D[j,*]$ is a new pivot and $\lp$ is
incremented.
\item\label{step:add-waiting}
If $j \neq \lp +1$, $D[j,*]$ is added to $W$ keeping track that
$\lp$ is the index of the last row $D[j,*]$ is already reduced with.
\end{enumerate}

Naturally, the above description leaves some freedom for the decision which row
to fetch and reduce next in Step~\ref{step:fetch}. We found
the following choice to be the most efficient for a wide range of examples:
When a thread fetches a row to be further reduced it prefers a row that
was already previously reduced. This often leads to a faster recognition of new
known pivots in Step~\ref{step:add-piv}.
Synchronization is needed in Steps~\ref{step:add-piv}
and~\ref{step:add-waiting}, besides this the threads can work independent of
each other. We handle the communication between the threads using spin locks
whose implementation w.r.t. a given used different parallel scheduler (see
Section~\ref{sec:schedulers}) might differ slightly.

Talking about load balancing it can happen that one thread gets
stuck in reducing already earlier reduced rows further, whereas other threads
fetch pristine rows and fill up $W$ more and more. In order to avoid this we use
the following techniques:
\begin{itemize}
\item If a thread has just fully reduced a row $r$ and thus adds a new known pivot,
this thread prefers to take an already reduced row from $W$ possibly waiting for $r$ to become a known pivot.
\item If a thread has added $t$ new rows to $W$ consecutively, it is triggered to
further reduce elements from $W$ instead of starting with until now untouched
rows from $D$.
\end{itemize}

For efficiency reasons we do not directly start with the discussed parallel
elimination, but we do a sequential elimination on the first $k$ rows \resp
multiline rows of $D$. In this way we can avoid high increasing on the waiting list
$W$ at the beginning, which would lead to tasks too small to benefit from the
available number of cores executing in parallel. Thus $k$ depends on the number
of threads used, in practice we found that $k = 2 \times (\text{number of
threads})$ is a good choice. Clearly, the efficiency of this choice depends on
how many of the first $k$ rows \resp multiline rows of $D$ reduce to zero in
this step. This is not a problem for full rank matrices coming from \Ffive \gb
computations.

\subsection{Different parallel schedulers}
\label{sec:schedulers}
We did some research on which parallel schedulers to be used in our library. For
this we tested not only well known schedulers like \openmp~\cite{openmp} and \tbb~\cite{tbb} but also
\kaapi~\cite{kaapi} and \starpu~\cite{starpu}. We also did experiments with \pthreads and own
implementations for scheduling. Most of the schedulers have advantages and
disadvantages in different situations like depending on sparsity, blocksizes or relying
on locking for the structured Gaussian Elimination. Moreover, all those packages
are actively developed and further improved, thus we realized different
behavoiour for different versions of the same scheduler. In the end we decided
to choose \openmp for the current state of the library.
\begin{enumerate}
\item It is in different situations usually not the fastest scheduler, but often
tends to be the fastest for the overall computation.
\item Our library should be plain C as much as possible, thus we discarded the
usage of \tbb which is based on high-level C++ features for optimal usage.
\item Current versions of \kaapi are able to interpret \openmp pragmas. Thus one can
use our library together with \kaapi by changing the linker call: instead
of {\tt libgomp} one has to link against {\tt libkomp} (see also
    Section~\ref{sec:er}).
\item Using \pthreads natively is error-prone and leads to code that is not
portable (it is not trivial to get them work on
Windows machines). \openmp's locking mechanism boils down
to \pthreads on UNIX and their pendants on Windows without having to deal
with different implementations.
\item \starpu's performance depends highly on the used data structures. Since the
representation of our data is special (see Sections~\ref{s:fmt} and~\ref{s:seq})
we need further investigations on how to get data and scheduler playing together
efficiently. Moreover, the fact that \starpu can be used for task scheduling even
on heterogeneous multicore architectures like CPU/GPU combinations makes it a good candidate
for further experiments.
\end{enumerate}

\section{Experimental results}
\label{sec:er}
\begin{table}[h]
	\centering
	\renewcommand{\tabcolsep}{1.1mm}
  \scalebox{0.7}{
	\begin{tabular}{ccc|cccc}
    \toprule
		\multicolumn{3}{c|}{\multirow{2}{*}{Matrix}} & Rows  & Columns & Nonzeros &
    Density\\
		& & & $\times 10^3$  & $\times 10^3$ & $\times 10^6$ & \%\\
    \midrule
    \texttt{F4} & \texttt{kat12} & \texttt{mat9} & 18.8 & 22.3 & 17.1 & 4.07 \\[1.1mm]
          & \texttt{kat13} & \texttt{mat2} & 4.68 & 6.53 & 1.45 & 4.74 \\
          &       & \texttt{mat3} & 12.1 & 14.6 & 7.78 & 4.35\\
          &       & \texttt{mat5} & 35.4 & 38.2 & 63.7 & 4.13 \\
          &       & \texttt{mat9} & 43.5 & 49.2 & 75.3 & 3.52\\[1.1mm]
          & \texttt{kat14} & \texttt{mat8} & 100  & 103  & 352 & 3.39 \\[1.1mm]
          & \texttt{kat15} & \texttt{mat7} & 168  & 178  & 832 & 2.77\\
          &       & \texttt{mat8} & 197  & 210  & 1,060 & 2.55 \\
          &       & \texttt{mat9} & 228  & 234  & 1,521 & 2.84\\[1.1mm]
          & \texttt{eco14}  & \texttt{mat24} & 105 & 107 & 91.1 & 0.81\\[1.1mm]
          & \texttt{eco16} & \texttt{mat13} & 157 & 141 & 293 & 1.32\\[1.1mm]
          & \texttt{mr-9-8-8-5} & \texttt{mat7} & 26.0 & 34.1 & 236 & 26.6\\
          &       & \texttt{mat8} & 22.1 & 34.6 & 189 & 25.5\\[1.1mm]
          & \texttt{rand16-d2-2} & \texttt{mat5} & 67.1  & 106  & 199 & 2.80\\
          & & \texttt{mat6} & 146  & 217  & 689 & 2.19\\[1.1mm]
          & \texttt{rand16-d2-3} & \texttt{mat8} & 587  & 874  & 4,328 & 0.84 \\
          &       & \texttt{mat9} & 980  & 1,428  & 8,378 & 0.60\\
          &       & \texttt{mat10} & 1,544  & 2,199  & 14,440 & 0.43\\
          &       & \texttt{mat11}& 2,287  & 3,226  & 23,823 & 0.32\\[1.1mm]
          & \texttt{rand18-d2-9} & \texttt{mat5} & 430 & 1,028 & 1,048 & 0.24 \\
          &  & \texttt{mat6} & 1,212 & 2,674  & 3,879 &
          0.12\\
    \cmidrule(){1-7}
    \texttt{F5} & \texttt{kat13} & \texttt{mat5}  & 28.4 & 35.5 & 26.8 & 2.66 \\
          &       & \texttt{mat6}  & 34.5 & 42.3 & 35.9 & 2.46 \\[1.1mm]
          & \texttt{kat14} & \texttt{mat7}  & 69.6 & 84.5 & 118 & 2.01\\
          &       & \texttt{mat8}  & 81.0 & 96.9 & 156 & 1.98 \\[1.1mm]
          & \texttt{kat15} & \texttt{mat7}  & 139  & 167  & 383 & 1.63 \\
          &       & \texttt{mat8}  & 168  & 199  & 507 & 1.51\\
          &       & \texttt{mat9}  & 187  & 219  & 640 & 1.56\\
          &       & \texttt{mat10} & 195  & 227  & 725  & 1.63\\[1.1mm]
          & \texttt{kat16} & \texttt{mat5}  & 83.8 & 110  & 139 & 1.50\\
          &       & \texttt{mat6}  & 168  & 208  & 485 & 1.38\\
          &       & \texttt{mat9}  & 393  & 456  & 2,234 & 1.25\\[1.1mm]
          & \texttt{cyc10} & \texttt{mat19}  & 192 & 256 & 1,182 & 2.40\\
          & & \texttt{mat20}  & 303 & 378 & 2,239 & 1.95\\[1.1mm]
          & \texttt{cyc10-sym1} & \texttt{mat17}  & 29.8 & 43.3 & 114 & 0.09\\[1.1mm]
          & \texttt{mr-9-10-7}  & \texttt{mat3}  & 20.1 & 74.5 & 532 & 35.5 \\
          & & \texttt{mat7}  & 88.5 & 192 & 4,055 & 23.8 \\[1.1mm]
          &  \texttt{rand16-d2-2} & \texttt{mat11} & 1,368 & 1,856  & 15,134 & 0.60\\
          &       & \texttt{mat12} & 1,806 & 2,425 & 22,385 & 0.51\\
          &       & \texttt{mat13}& 2,310 & 3,076 & 31,247 & 0.44\\[1.1mm]
          & \texttt{rand16-d2-3} & \texttt{mat8} & 578 & 871  & 3,140 & 0.62\\
          &  & \texttt{mat9} & 973 & 1,426  & 6,839 & 0.49\\
          &       & \texttt{mat10} & 1,532 & 2,198 & 13,222 & 0.39\\
          &       & \texttt{mat11}& 2,286 & 3,226 & 23,221 & 0.31\\
          &       & \texttt{mat12}& 3.266 & 4,550 & 37,796 & 0.25\\[1.1mm]
          & \texttt{rand18-d2-9} & \texttt{mat5} & 429 & 1,027  & 9,65 &
          0.22\\
          & & \texttt{mat7} & 3,096 & 6,414  & 12,594 &
          0.06\\
          &  & \texttt{mat11} & 1,368 & 1,856  & 15,135 & 0.60\\
    \bottomrule
	\end{tabular}
  }
	\caption{Some matrix characteristics}
	\label{tab:expe:carac}
\end{table}
The following experiments were performed on \url{http://hpac.imag.fr/} which
is a NUMA architecture of 4x8 processors. Each of the 32 non hyper-threaded
Intel(R) Xeon(R) CPUs cores clocks at 2.20GHz (maximal turbo frequency on single core 2.60GHz).
Each of the 4 nodes has 96Gb of memory, so we have 384Gb of RAM in total.
The compiler is \texttt{gcc-4.9.2}. The timings do not include
the time spent on reading the files from disk. We state matrix characteristics
of our example set in Table~\ref{tab:expe:carac}.

We use various example sets: There are well known benchmarks like \texttt{Katsura},
\texttt{Eco} and \texttt{Cyclic}\footnote{Also including a version where we have
applied the symmetry of the cyclic group action of degree $1$,
see~\cite{faugere-svartz-2013}.}. 
Moreover, we use matrices from minrank problems arising in cryptography.
Furthermore we have random dense systems \texttt{randx-d2-y-mat*} in \texttt{x} variables, all input
polynomials are of degree $2$. Then we deleted \texttt{y} polynomials
to achieve higher-dimensional benchmarks.
All examples are done over the biggest $16$-bit prime field,
$\mathbb{F}_{65521}$. We use the \texttt{uint16} coefficient representation in
\gbla. If not otherwise stated \gbla's timings are done using \openmp as parallel
scheduler.


\subsection{Behaviour on \Ffive matrices}
We show in Table~\ref{tab:expe:fl} a comparison with Faug\`ere and Lachartre's
\fli from \cite{FaLa10} and \gbla. Timings are in seconds, using 1, 16 or 32
threads. This is done for \Ffive matrices, thus we can use \gbla's new
order of operations (see Section~\ref{sec:new-order}) to compute a Echelon form
and to verify that the matrices have full rank.
\def\mrm{$\text{minrank}_{9,10,7}$}
\def\pmrm{\phantom{\mrm}}

\begin{table}
	\centering
	\renewcommand{\tabcolsep}{1mm}
  \scalebox{0.72}{
	\begin{tabular}{c|ccc|ccc|ccc}
		\toprule
    Implementation & \multicolumn{3}{c|}{\fli} &
    \multicolumn{3}{c|}{\gblaone} & \multicolumn{3}{c}{\gblatwo}\\
    \midrule
		\Ffive Matrix / \# Threads& 1 & 16 & 32  & 1 & 16 & 32 & 1 & 16 & 32 \\
    \midrule
\texttt{kat13-mat5} & 16.7& 2.7 & 2.3 &  \ok{14.5} & 2.02 & 1.87
    & \ok{14.5} & \ok{1.73} & \ok{1.61}\\
\texttt{kat13-mat6} & 27.3 &  4.15  & 4.0 & \ok{ 23.9} & 3.08 &
  2.65 & 25.9 & \ok{3.03} & \ok{2.28} \\[1.1mm]
\texttt{kat14-mat7} & 139 & 17.4 & 16.6 & 142 & 13.4 & 10.6
  & \ok{122} & \ok{11.2} & \ok{8.64}\\ 
\texttt{kat14-mat8} & 181 & 24.95 & 23.1 & 177 & 16.9 &
  12.7 & \ok{158} & \ok{14.7} & \ok{10.5} \\[1.1mm]
\texttt{kat15-mat7} & 629 & 61.8 & 55.6 & 633 & 55.1 & 38.2
& \ok{553} & \ok{46.3} & \ok{30.7}\\[1.1mm]
\texttt{kat16-mat6} & 1,203 & 110 & 83.3 & 1,147& 98.7 &
  69.9 & \ok{988}& \ok{73.9} & \ok{49.0}\\[1.1mm] 
\texttt{mr-9-10-7-mat3} & \ok{ 591} & 70.8 & 71.3 & 733 & 57.3 &
  37.9 & 747 & \ok{52.8} & \ok{33.2} \\
\texttt{mr-9-10-7-mat7} & 15,787 & 1,632 & 1,565 & \ok{15,416} & 1,103 & 793
   & 15,602 & \ok{1,057} & \ok{591}\\[1.1mm]
\texttt{cyc10-mat19} & 7,482 & 693 & 492 & 1,291 & 135 & 103 & \ok{1,030} & \ok{80.3} &
\ok{62.9}\\
\texttt{cyc10-mat20} & 17,853  & 1,644 & 1,180 & 2,589 & 274 & 209 & \ok{2,074} & \ok{171} & \ok{152}\\
\texttt{cyc10-sym1-mat17} & 11,083 & 1,982 & 1,705 & 2,463 & 465 & 405 & \ok{2,391} & \ok{275} & \ok{245}\\[1.1mm]
    \texttt{rand16-d2-2-mat11} & mem & mem & mem & \ok{2,568} & 946 & 883 &
    4,553 & \ok{425} & \ok{360}\\
    \texttt{rand16-d2-2-mat12} & mem & mem & mem & \ok{5,751} & 1,252 & 1,219 & 6,758 & \ok{632} & \ok{527}\\
    \texttt{rand16-d2-2-mat13} & mem & mem & mem & mem & mem & mem & \ok{8,435} & \ok{816} & \ok{721}\\[1.1mm]
    \texttt{rand16-d2-3-mat8} & 2,084 & 500 & 472 & 2,243 & 339 & 282 &
    \ok{1,654} & \ok{144} & \ok{106}\\
    \texttt{rand16-d2-3-mat9} & bug & bug & bug & 2,938 & 827 & 781 & \ok{2,308}
    & \ok{236} & \ok{227}\\
    \texttt{rand16-d2-3-mat10} & mem & mem & mem & \ok{2,528} & 922 & 940 & 4,518 &
    \ok{427} & \ok{372}\\
    \texttt{rand16-d2-3-mat11} & mem & mem & mem & mem & mem & mem & \ok{11,254} & \ok{931}
    & \ok{696}\\
    \texttt{rand16-d2-3-mat12} & mem & mem & mem & mem & mem & mem & \ok{15,817}&
    \ok{1,369} & \ok{1,150}\\[1.1mm]
    \texttt{rand18-d2-9-mat5} & 1,469 & 287 & 250 & 350 & 297 & 306 & \ok{340}
    & \ok{52.9} & \ok{50.3}\\
    \texttt{rand18-d2-9-mat7} & mem & mem & mem & mem & mem & mem & \ok{8,752}
    & \ok{1,112} & \ok{1,098}\\
    \texttt{rand18-d2-9-mat11} & bug & bug & bug & \ok{2,540} & 923 & 882 &
    4,600 & \ok{415} & \ok{363}\\
    \bottomrule
	\end{tabular}
  }
	\caption{\flr \vs{} \gbla (time in seconds)}
	\label{tab:expe:fl}
\end{table}

Usually \gblaone is faster on one core than \fli, \gblatwo is even faster
than \gblaone. Both \gbla implementations have a much better scaling than \fli,
where \gblatwo preforms better than \gblaone, even scaling rather good for
smaller examples, where the overhead of scheduling different threads starts to
become a bottleneck. The only example where \fli is faster than \gbla is
\texttt{mr-9-10-7-mat3}, a very dense ($35.5\%$) matrix. This good behaviour for
\fli might be triggered from the fact that \fli allocates all the memory needed
for the computation in advance. Usually the user does not know how much memory the
computation might need, so this approach is a bit error-prone. Still, \fli is
faster than \gbla only on one core, starting to use several CPU cores the better
scaling of \gbla wins (already at $2$ cores the timings are nearly identical).
Moreover, for dense matrices like the minrank ones we can see a benefit of the
multiline structure, at least for fewer cores. Once the number of cores
increases the better scaling of \gblatwo is favourable.

For \texttt{cyc10-sym1-mat17} the speedup between
$16$ and $32$ is quite small. Due to the applied symmetry the matrix is already
nearly reduced, so the scheduling overhead has a higher impact
than the gain during reduction for anything greater than $16$
cores.

For the higher-dimensional random examples the row dimension of
$C$ and $D$ is very small ($<300$). Our new order of
operations (Section~\ref{sec:new-order}) enables \gbla to reduce matrices the \fli
is not able to handle.
Even \gblatwo reaches for \texttt{rand18-d2-9-mat7}
the memory limit of the machine, but it is still able to reduce the
matrix. Memory overhead due to multilines hinders \gblaone to
compute \texttt{rand16-d2-3-mat11}, but is more efficient on
\texttt{rand16-d2-3-mat10} for one core.

\subsection{Behaviour on \Ffour matrices}
In Table~\ref{tab:expe:gbla} we compare \magma-2.19~\cite{bcpMagma} to \gbla. Since there is no
\Ffive implementation in \magma we can only compare matrices
coming from \Ffour computations. Since \magma is closed source we are not able
to access the specialized linear algebra for \gb computations directly. Thus,
we are comparing the same problem sets with the same degrees running \magma's
\Ffour implementation. Note that \magma generates matrices that are, for the
same problem and same degree, slightly larger, usually $5$ to $10$\%. Note that
we use only \magma's CPU implementation of \Ffour, but not the rather new GPU
one. We think that it is not really useful to compare GPU and CPU parallelized
code. Furthermore, most of our examples are too big to fit into the RAM of a
GPU, so data copying between CPU and GPU might be problematic for an accurate comparison.

\begin{table}
	\centering
	\renewcommand{\tabcolsep}{1mm}
  \scalebox{0.725}{
	\begin{tabular}{c|c|ccc|ccc}
    \toprule
		Implementation & \magma & \multicolumn{3}{c}{\gblaone} &
    \multicolumn{3}{|c}{\gblatwo} \\
    \midrule
		\Ffour Matrix / \# Threads & 1 & 1 & 16  & 32 & 1 & 16  & 32 \\
    \midrule
    \texttt{kat12-mat9} & \ok{ 11.2} & 11.4       & 1.46  & 1.60 & 11.3 &
    1.40 & 1.40\\[1.1mm]
    \texttt{kat13-mat2} & \ok{ 0.94}& 1.18       & 0.38  & 0.61 & 1.11 & 0.26
    & 0.33\\
    \texttt{kat13-mat3} & 9.33 & 11.0       & 1.70  & 3.10 & \ok{8.51} & 1.07
    & 1.13\\
    \texttt{kat13-mat9} & 168 & 165    & 16.0  & 11.8 & \ok{114} & 9.74 & 6.83 \\[1.1mm]
    \texttt{kat14-mat8} & 2,747  & 2,545   & 207   & 165 & \ok{1,338} & 104 &
    65.8\\[1.1mm]
    \texttt{kat15-mat7} & 10,345 & 9,514  & 742   & 537 & \ok{4,198} & 298 & 195 \\
    \texttt{kat15-mat8} & 13,936 & 12,547 & 961   & 604  & \ok{6,508} & 470 & 283 \\
    \texttt{kat15-mat9} & 24,393 & 22,247 & 1,709 & 1,256 & \ok{10,923} & 779 & 450\\[1.1mm]
    \texttt{eco14-mat24} & 524 & 169 & 22.2 & 21.9 & \ok{146} & 16.2 & 16.5\\[1.1mm]
    \texttt{eco16-mat13} & 6,239 & 1,537 & 184 & 176 & \ok{1,346} & 104 & 72.9 \\[1.1mm]
    \texttt{mr-9-8-8-5-mat7} & 1,073 & 1,080 & 88.5 & 57.9 & \ok{550} & 41.6 & 24.7\\
    \texttt{mr-9-8-8-5-mat8} & 454  & 600 & 48.5 & 30.3 & \ok{318} & 25.6 & 14.9\\[1.1mm]
    \texttt{rand16-d2-2-mat5} & 740 & 778 & 62.2 & 40.8 & \ok{589} & 43.6 & 28.6  \\
    \texttt{rand16-d2-2-mat6} & 4,083 & 4,092 & 375 & 219  & \ok{3,054} & 224 & 133\\[1.1mm]
    \texttt{rand16-d2-3-mat8} & 55,439  & 48,008 & 3,473 & 2,119 & \ok{26,533} & 1,782 & 1,027 \\
    \texttt{rand16-d2-3-mat9} & 91,595 & 65,126 & 4,869 & 2,983 & \ok{39,108} &
    2,614 & 1,372 \\
    \texttt{rand16-d2-3-mat10} & mem  & - & 9,691 & 6,223 & - & 3,820 & 1,972 \\
    \texttt{rand16-d2-3-mat11} & mem & mem & mem &mem & - & 5,399 & 2,385 \\[1.1mm]
    \texttt{rand18-d2-9-mat5} & 2,020 & 1,892 & 414 & 388 & \ok{630} & 63.1  & 61.8  \\
    \texttt{rand18-d2-9-mat6} & 4,915 & 6,120 & 981  & 941 & \ok{1,736} & 220 &
    218\\
    \bottomrule
	\end{tabular}
  }
	\caption{\magma \vs{} \gbla (time in seconds)}
	\label{tab:expe:gbla}
\end{table}

For small examples \magma, not splicing the matrices, has an advantage. But
already for examples in the range of $10$ seconds \gbla, especially \gblatwo
gets faster on single core. The difference between \magma and \gblaone is rather
small, whereas \gblatwo becomes more than twice as fast. Moreover, \gblaone and
\gblatwo scale very well on $16$ and $32$ cores. Due to lack of space we do not
state timings for the \fli. It behaves in nearly all examples like expected: Due
to preallocation of all memory it is very fast on sequential computations
(nearly as fast as \gblatwo), but it scales rather bad. For example, for
\texttt{kat14-mat8} \fli runs in $1,571$s, $861$s and $868$s for $1$, $16$ and
$32$ cores, respectively. Also note that \fli's memory consumption is higher
than \gbla's.

For the random, higher
dimensional examples \magma cannot reduce matrices starting from \texttt{rand16-d2-3-mat9}
due to the \texttt{float} representation of the
matrix entries and resulting higher memory usage on the given machine.
For \texttt{rand16-d2-3-mat11} even
\gblaone consumes too much memory by using multilines and thus introducing too many
zeros (see Section~\ref{sec:multiline}). Even \gblatwo comes to
the limit of our chosen compute server, but it can still reduce the matrix: At
the end of the computation the process consumed $98\%$ of the machine's RAM.

\subsection{Comparing \openmp (\omp) and \kaapi (\xk)}
We compare the different behaviour of the parallel schedulers that can be used
in \gbla (see also Section~\ref{sec:schedulers}): The default scheduler in \gbla
is \omp, here we use the latest stable version $4.0$. \xk can interpret
\omp pragmas, too, so we are able to run \gbla with \xk by just changing
the linker call from \texttt{libgomp} to \texttt{libkomp}. The latest stable
version of \xk we use is $3.0$.
In Table~\ref{tab:gbla-omp-kaapi} we compare both schedulers on
representative benchmarks in \gblaone and \gblatwo on $16$ and $32$ cores.
The timings show that in many examples both schedulers are on par. \xk tends
to be a bit more efficient on $32$ cores, but that is not always the case.
\texttt{F4-kat15-mat8} and \texttt{F4-kat15-mat9} are cases where \xk has
problems on $32$ cores for \gblaone. This comes from the last step, the
structured Gaussian Elimination of $D$ where \gblaone, using
multilines, cannot sort the pivots which seems to become a bottleneck for \xk's
scheduling. For the same examples
in \gblatwo (now with sorting of pivots) we see that \xk is even a bit faster
than \omp. 
All in all, in our setting both schedulers behave nearly equal.

\begin{table}
	\centering
	\renewcommand{\tabcolsep}{0.7mm}
  \scalebox{0.75}{
	\begin{tabular}{c|cccc|cccc}
    \toprule
		Implementation & \multicolumn{4}{c|}{\gblaone} &
     \multicolumn{4}{c}{\gblatwo} \\
    \midrule
		\# Threads & \multicolumn{2}{c}{16} &
    \multicolumn{2}{c|}{32} & \multicolumn{2}{c}{16} &
    \multicolumn{2}{c}{32}\\
    Matrix / Scheduler& \footnotesize{\omp} & \footnotesize{\xk}& \footnotesize{\omp} &
    \footnotesize{\xk}& \footnotesize{\omp} & \footnotesize{\xk}&
    \footnotesize{\omp} & \footnotesize{\xk}\\
    \midrule
    \texttt{F4-kat15-mat8} & 961   & \ok{916} & \ok{604} & 1,223 & 470 &
    \ok{463}  & 283 & \ok{277} \\
    \texttt{F4-kat15-mat9} & 1,709 & \ok{1,679} & \ok{1,256} & 2,122 & 779 &
    \ok{774} & 450 & \ok{431}\\
    \texttt{F4-rand16-d2-3-mat8} & 3,4732 & \ok{3,447} & 2,119 & \ok{1,964} &
    \ok{1,782} & 1,818 & 1,027 & \ok{1,017} \\
    \texttt{F4-rand16-d2-3-mat9} & \ok{6,956} & 7,073 & 4,470 & \ok{3,783} &
    3,214 & \ok{3,141} & \ok{1,776} & 1,785 \\
    \cmidrule(){1-9}
    \texttt{F5-kat16-mat6} & \ok{98.7} & 105 & 69.9 & \ok{67.2} & \ok{73.9} &
      75.3 & \ok{49.0} & \ok{49.0}\\
    \texttt{F5-mr-9-10-7-mat3} & \ok{57.3} & 59.6 & \ok{37.9} & 38.8 & \ok{52.8} & 54.8 & \ok{33.2} & 34.7 \\
    \texttt{F5-cyc10-mat19} & \ok{135} & 140 & 103 & \ok{101} & \ok{80.3} & 86.7 &
    62.9 & \ok{60.9}\\
    \texttt{F5-cyc10-mat20} & \ok{274} & 292 & 209 & \ok{206} & \ok{171} & 203 & 152 & \ok{141}\\
    \texttt{F5-cyc10-sym1-mat17} & \ok{465} & 496 & \ok{405} & 406 & 275 &
    \ok{272} & 245
    & \ok{217}\\
    \bottomrule
	\end{tabular}
  }
	\caption{\openmp \vs{} \kaapi (time in seconds)}
	\label{tab:gbla-omp-kaapi}
\end{table}


%


\section{Conclusion}\label{s:conclusion}
We presented the first
open-source, plain C library
for linear algebra specialized for matrices coming from \gb computations
including various new ideas exploiting underlying structures. This led to more
efficient ways of performing the \flr and improved parallel scaling. Moreover,
the library uses a new compressed file format that enables us to generate
matrices not feasible beforehand. Corresponding routines for dumping and
converting own matrices are included  such that researchers are able to use
their own data in our new format in \gbla.

Also the time needed to reduce $D$ during  \flr is in general very
small compared to the overall reduction, we
plan to investigate our parallel structured Gaussian elimination implementation
in the future. For this we may again copy $D'$ first to a different data
representation and use external libraries for fast exact linear algebra such as
\fflas~\cite{DGPS14} in given situations.

\bibliographystyle{abbrv}

\end{document}